\def\tf{t_s(r)}

\documentclass[oneside,a4paper,10pt]{amsart}
\usepackage{amsmath}               
\usepackage{amscd}
\usepackage{amsthm}                
\usepackage{times}

\numberwithin{equation}{section}

\title[Sufficient condition for Blackhole formation in spherical
gravitational collapse ]%
{Sufficient condition for Blackhole formation in spherical
gravitational collapse}
\author[R.\ Giamb\'o ,\ F.\ Giannoni]{Roberto Giamb\'o, Fabio Giannoni}
\address{Dipartimento di Matematica e Fisica\hfill\break\indent
Universit\`a di Camerino, Italy}
\email{giambo@campus.unicam.it , giannoni@campus.unicam.it}
\author[G.\ Magli]{Giulio Magli}
\address{Dipartimento di Matematica,\hfill\break\indent Politecnico di
Milano, Italy}
\email{magli@mate.polimi.it}
\begin{document}
\swapnumbers
\theoremstyle{plain}\newtheorem{teo}{Theorem}[section]
\theoremstyle{plain}\newtheorem{prop}[teo]{Proposition}
\theoremstyle{plain}\newtheorem{lem}[teo]{Lemma}
\theoremstyle{plain}\newtheorem{cor}[teo]{Corollary}
\theoremstyle{definition}\newtheorem{defin}[teo]{Definition}
\theoremstyle{remark}\newtheorem{rem}[teo]{Remark}
\theoremstyle{plain} \newtheorem{assum}[teo]{Assumption}
\theoremstyle{definition}\newtheorem{example}[teo]{Example}

\begin{abstract}

A sufficient condition for the validity of Cosmic Censorship
in spherical gravitational collapse is formulated and proved.
The condition relies on an attractive mathematical property
of the apparent horizon, which holds if ``minimal''
requirements of physical reasonableness are satisfied
by the matter model.

\end{abstract}

\maketitle

\begin{section}{Introduction}\label{sec:intro}

Predicting the final state (blackhole or naked singularity) of the
gravitational collapse of an initially regular  matter
distribution in General Relativity is still, to a large extent, an
open problem, even in the special case of spherical symmetry.
In fact, Penrose's ``Cosmic Censorship'' conjecture \cite{p}
stating that "every physically reasonable collapsing matter
distribution forms a blackhole"
seems to be overruled by several
examples, in which what is likely to be a physically reasonable
distribution of matter forms a visible null singularity (for
instance in the dust and in the Vaydia case, see \cite{jo} and references
therein)
or even a timelike singularity, as in the case of gravitating clusters
of
rotating particles \cite{gm1,h1,h2,gmj}. It is, therefore, necessary to
obtain a
rigorous formulation of censorship in mathematical terms, in order
to be able to translate the conjecture in mathematically
demonstrable assertions. In this formulation, the physics of the problem
is obviously
expected to play a fundamental role. For instance, the weak energy
condition has naturally
to be expected as one of the hypotheses of any cosmic censorship
theorem.

In the present letter, we report  on an
investigation carried out recently devoted to the application of
tecniques of non-linear o.d.e. to the censorship problem in
spherical symmetry. The main result is a non-existence theorem for
geodesics meeting the singularity in the past, which gives a
sufficient condition for a spherically symmetric non-static metric
to represent blackhole formation. The condition can be used as a
test bed for probing models of collapse without knowing the
explicit solution of the Einstein field equations.

\end{section} 

\begin{section}{The analytical framework in brief}\label{sec:dust}

We consider a spherically symmetric collapsing object in full
generality. The matter model can thus be any model compatible with
the weak energy condition.
The general, spherically symmetric,
non--static line element in comoving coordinates
$t,r,\theta,\varphi$ can be written in terms of three functions
$\nu , \lambda , R$ of $r$ and $t$ only as follows:
$$
\mathrm ds^2= -e^{2\nu}\mathrm dt^2 + e^{2\lambda}\mathrm dr^2 + R^2
(\mathrm d\theta^2 +
\sin^2\theta\mathrm d\varphi^2)\ ,
$$
A fundamental quantity is the mass function $m(r,t)$ defined
through
\begin{equation}\label{t}
1-\frac{2m}R=g^{\mu\nu}\frac{\partial R}{\partial
x^\mu}\frac{\partial R}{\partial x^\nu}
\end{equation}
that is
\begin{equation}\label{m}
m(r,t)=\frac R2 \left[1-(R')^2e^{-2\lambda}+(\dot
R)^2e^{-2\nu}\right]\
\end{equation}
where a dash and a dot denote derivatives with respect to $r$ and
$t$ respectively. Field equations $G^0_0=8\pi T^0_0$ and
$G^r_r=8\pi T^r_r$ relate derivatives of $m$ to the energy density
$\epsilon$ and
the radial stress $p_r$ of the material:

\begin{equation}\label{g1}
m'  =4\pi \epsilon R^2 R'\ ,
\end{equation}

\begin{equation}\label{eq:G2}
\dot m =-4\pi p_r R^2 \dot R\ ,
\end{equation}

We consider only matter configurations admitting a regular center,
and we suppose always that the collapse starts from regular
initial data on a Cauchy surface ($t=0$, say), so that the
singularities forming are a genuine outcome of the dynamics. If
the solution is initially regular, \eqref{g1} implies that the
energy density - and therefore the spacetime - becomes singular
whenever $R$ or $R'$ vanish during the evolution. It follows, that
singularities can be of two different kinds: {\it shell crossing},
at which $R'$ vanishes while $R$ is non-zero, and {\it shell
focusing} at which $R$ vanishes. The shell crossing singularities
have been frequently considered as ''weak'' although no proof of
extensibility is as yet available in the literature. In any case,
in most physically interesting situations such singularities do
not occur, so that we shall concentrate attention here only on the
shell focusing case. Therefore, we assume that $R'$ is initially
non zero for non-zero $r$ and remains positive up to the formation
of the focussing singularities.

The locus of the zeroes of the function $R(r,t)$ defines the {\it
singularity curve} $\tf $ by the relation $R(r,\tf )=0$.
Physically, $\tf$ is the comoving time at which the shell of
matter labeled by $r$ becomes singular. The singularity forming at
$r=0,t=t_s (0)$ is called {\sl central} as opposed to those
occurring at $r=r_0>0,t=t_s (r_0)$.

A singularity cannot be naked if it occurs after the formation of
the apparent horizon. The apparent horizon ($t_h(r)$, say) is the
boundary of the region of trapped surfaces and it is therefore
defined, due to equation \eqref{t}, by $R(r,t_h(r))=2m(r,t_h(r))$.
Since $R$ vanishes at a singularity, any naked singularity in
spherical symmetry must be massless; at a massless singularity the
horizon and the singularity form simultaneously. Since regularity
of the center up to singularity formation requires
$m(0,t)=0\,\forall t< t_s(0)$, the center is always a candidate
for nakedness. Other points ("non-central points") of the
singularity curve can be naked only in presence of radial
tensions. In fact, due to eq. \eqref{eq:G2}, the mass is
increasing in time in a collapsing ($\dot R<0$) situation, if the
radial pressure is positive \cite{cj}.

To analyze the causal structure, observe that, if the singularity
is visible to nearby observers, at least one outgoing null
geodesic must exist, that meets the singularity in the past. Such
a geodesic will be a solution of
\begin{equation}\label{f}
\frac{\mathrm dt(r)}{\mathrm dr}=\varphi (r,t)
\end{equation}
where
\begin{equation}\label{eq:phi}
\varphi(r,t):=\sqrt{ -\frac{g_{rr}}{g_{00} } }=e^{\lambda-\nu}
\end{equation}
with initial datum $t(r_0)=t_s(r_0)$. For a problem of this kind, in
which the initial point is singular (the function $\varphi$ is not
defined at $(r_0,t_s(r_0))$) no general results of existence/non
existence are known. As a consequence, in the literature, an
approach has been developed (see e.g. \cite{jo}) which makes use
of l'Hospital theorem to identify the possible values of the
tangent of the geodesic curve at the singularity. This is the
approach that allowed a full understanding of the dust and of the
Einstein cluster cases recalled above. However, to be successful,
this tecnique requires complete integration of the field
equations, a result which is far beyond our present understanding
even in the simple case of the barotropic perfect fluid. On the
other side, what actually enters the problem of the causal
structure is only the function $\varphi$, not the whole solution
of the field equations. Therefore, one can consider this problem
as an existence/non-existence problem for the non-linear o.d.e.
\eqref{f}, in which the mathematical structure of the Einstein
equations as well as the physics of the problem (like e.g.
formation of trapped surfaces, weak energy condition) play a
fundamental role.

\end{section}

\begin{section}{The non-existence theorem}\label{sec:non-ex}

From now on we will assume the following condition on
$p_r,\,\epsilon$ (see Remark \ref{rem:wec} below):
\begin{equation}\label{C}
\epsilon>0 \ \ , \ \ p_r \geq {\rm Max}\{ -\epsilon ,
-\frac{1}{8\pi R^2}\}.
\end{equation}

\begin{teo}\label{teo:non-ex1}
If \eqref{C} holds and $\frac{\partial \varphi}{\partial t}\le 0$ in a neighborhood of
$(r_0,t_s(r_0))$, with $r_0\ge 0$, the singularity forming at $(r_0,t_s(r_0))$ is
covered.
\end{teo}

The theorem is based on a remarkable mathematical property of the
apparent horizon, which we address in the following

\begin{lem}\label{lem:subsol}
If \eqref{C} holds
then there exists $r_*>r_0$
such that the apparent horizon $t_h(r)$ is a subsolution of
\eqref{f} for $r\in(r_0,r_*)$.
\end{lem}

\begin{proof}
Differentiating the equation $R(r,t_h(r))=2m(r,t_h(r))$ with respect to
$r$ we get
$$
\frac{\mathrm dt_h}{\mathrm dr}=-\frac{2m'-R'}{2\dot m-\dot R} =-\Gamma
\frac{R'}{\dot R}
$$
where
$$
\Gamma:=\frac{1-8\pi\epsilon R^2}{1+8\pi p_rR^2}
$$ and
the field equations \eqref{g1}--\eqref{eq:G2} have been used.
On the other end, from eq. \eqref{m} we get
$$
\varphi(r,t)=e^{\lambda -\nu}=-\left[1+e^{2\nu}\dot
R^{-2}\left(1-\frac{2m}R\right)\right]^{-\frac 12}\frac{R'}{\dot
R}
$$
where a (crucial) minus sign is due to the fact that we consider a
collapsing scenario (thus $\dot R$ is strictly negative, at least
near the singularity).
At $t=t_h$ the quantity in square brackets
is equal to one so that $\varphi(r,t_h)=-R'/\dot R$. It follows
$dt_h/dr \le\varphi (r,t_h)$ whenever the quantity $\Gamma$ is less
than or equal to unity, that is
\begin{equation}\label{eq:ineq}
\frac{8\pi R^2(\epsilon +p_r)}{8\pi p_rR^2+1}>0.
\end{equation}

\end{proof}

\begin{rem}\label{rem:wec}
If $8\pi p_rR^2+1$ is negative, the radial pressure would diverge to
minus infinity at a singularity,
a manifestly unphysical situation. Thus, we consider
further only matter models satisfying $p_r>-1/8\pi R^2$. This is a very
weak bound;
obviously it must be considered only if tensions are present, and at
most
it has to be extended to the whole of the collapsing object
(in this case it suffices to require $p_r>-1/8\pi R^2(r_b,t)$ where $r_b$
is the boundary
of the object).
Once this is satisfied, the inequality \eqref{eq:ineq} holds if $\epsilon +p_r >0$.
\end{rem}
\bigskip

{\sl Proof of Theorem \ref{teo:non-ex1}.}
Let $t_\rho(r)$ the solution of $t'(r)=\varphi(r,t(r))$ such that
$t_\rho(r_0)=t_s(r_0)$. By contradiction we suppose the existence
of $r_1>0$ such that $t_\rho(r_1)<t_h(r_1)$ and $t_\rho(r)\le
t_h(r),\,\forall r\in [r_0,r_1]$. We can suppose $r_1<r_*$, where
$r_*$ comes from Lemma \ref{lem:subsol}. Since
$t_\rho(r_0)=t_h(r_0)$, one has
\begin{multline}\label{eq:contra1}
0<t_h(r_1)-t_\rho(r_1)=\left(t_h(r_1)-t_\rho(r_1)\right)-\left((t_h(r_0)-t
_\rho(r_0))\right)=\\
\left(t'_h(\xi)-t'_\rho(\xi)\right)\,r_1=(t'_h(\xi)-\varphi(\xi,t_\rho(\xi
)))\,r_1,
\end{multline}
where $\xi\in(r_0,r_1)$. Using Lemma \ref{lem:subsol} it is
$t'_h(\xi)\le \varphi(\xi,t_h(\xi))$, and hence
\begin{equation}\label{eq:contra2}
0< t'_h(\xi)-\varphi(\xi,t_\rho(\xi))\le\frac{\partial\varphi}{\partial
t}(\xi,\theta)\,\,(t_h(\xi)-t_\rho(\xi)).
\end{equation}
Combining \eqref{eq:contra1} and \eqref{eq:contra2} one gets a
contradiction if $\frac{\partial\varphi}{\partial
t}(\xi,\theta)\le 0$.
\qed
\end{section}

\begin{section}{Discussion and conclusions}\label{sec:ex}

It is important to analyze the relationship of the condition
\eqref{C} of the former lemma with the weak energy condition ({\sl
wec}). The latter requires $\epsilon >0$, $\epsilon +p_r \geq 0$
and $\epsilon +p_t\geq 0$. Here, there is no condition on the
tangential pressure. If the radial pressure is positive, then
\eqref{C} coincides with the remaining inequalities of {\sl wec},
and is therefore {\it weaker}. In presence of tensions ($p_r<0$),
it places a lower bound on the radial stress which again coincides
with that coming from {\sl wec} unless $\epsilon>1/8\pi R^2$, when
the lower bound on $p_r$ must be explicitly required.

The condition stated is only sufficient. As a test-bed we can use
the dust case, in which the exact solution is known in closed form
and the nature of the singularities is known in full details
\cite{js} (a discussion of dust collapse in terms of existence/non
existence of solutions for non-linear o.d.e. can be found in
\cite{gm}).

The solution space for marginally bound collapse can be
parameterized in terms of an integer ($n$, say) giving the order
of the first non-vanishing derivative of the initial density
profile at the center (since $p_r$ is zero, the center is the unique point
that can be naked).
One easily finds that, for $n=1$ and $n=2$,
$\partial\varphi/\partial t$ is positive near the center, while
for $n>2$ one has
\begin{equation}\label{eq:dphidt(th)}
\frac{\partial\varphi}{\partial t}(r, t_h(r))=- \frac 1r
\left(1-\beta_n r^{n-3}\right)+O(r^{n-2})
\end{equation}
where $\beta_n$ is a positive quantity proportional to minus the
first non vanishing derivative of the density at the center. Thus,
the singularity is certainly covered for any $n>4$ and for $n=3$
if $\beta_3<1$. Actually, however, we know that the singularity is
naked if $n$ equals one, if $n$ equals two, or if $n$ equals three
but $\beta_3$ is greater than a positive value $\beta_c$ which in
turn is {\it greater} than one \cite{js} ( $\beta_c=(26 + 15\sqrt
3)/4$). Therefore, there exists is a region of the solutions space
in which blackholes still form, while our (thereby only
sufficient) condition does not hold. Interestingly enough, this
phenomenon occurs - at least in dust spacetimes - near to the
transition of the critical parameter. Physically, it reflects the
fact that the absence of apparent horizon formation prior to
singularity does not necessarily implies nakedness \cite{ff2}.

The results of the present paper
can be applied to all the matter
models that admit regular initial data in spherical symmetry. For
fluid sources, the space of all such solutions can be
parameterized in terms of three functions, namely the initial
density, the initial velocity, and the equation of state
\cite{bari}. A physically viable formulation of a Cosmic
Censorship theorem for fluid bodies would therefore rely in a
classification of this solution space in terms of the final
outcome of the collapse. In this context the theorem proved here
can be used to characterize a "large" subset of this space, that
contains only blackholes. Work in this direction is in progress.

Analyzing in full generality what happens at the "boundary" of
this subset, i.e. when a transition from blackholes to naked
singularities is expected to occur, looks, unfortunately, still a
quite far objective.

\end{section}


\begin{thebibliography}{99}



\bibitem{p} R. Penrose,
Nuovo Cimento {\bf 1} 252 (1969).

\bibitem{jo}  P. S. Joshi,
{\it Global aspects in gravitation and cosmology},
(Clarendon press, Oxford, 1993).

\bibitem{gm1} G. Magli,
Class. Quantum Grav. {\bf 15} 3215 (1998).

\bibitem{h1} T. Harada, H. Iguchi and K. Nakao,
Phys. Rev. D {\bf 58} R041502 (1998).


\bibitem{h2}
Kudoh, H. Harada, T. Iguchi, H. Phys. Rev. D (3) (2000) 104016,

\bibitem{gmj} S. Jhingan and G. Magli, Phys. Rev. {\bf D61} (2000)
124006

\bibitem{cj} F. I. Cooperstock, S. Jhingan, P. S. Joshi and T. P.
Singh,
Class. Quantum Grav. {\bf 14} 2195 (1997).


\bibitem{js} T. P. Singh and P. S. Joshi,
Class. Quantum Grav. {\bf 13} 559 (1996).


\bibitem{gm} R. Giambo' and G. Magli, preprint.


\bibitem{ff2} Jhingan, S. Joshi, P. S.; Singh, T. P.
Class.. Quantum Grav. 13 (1996) 3057.

\bibitem{bari} Jhingan, S. and Magli, G., In {\sl Recent
developments in General Relativity} B. Casciaro, D. Fortunato, A.
Masiello, M. Francaviglia ed., Springer Verlag (Berlin) (2000).


\end{thebibliography}
\end{document}